\documentclass[fleqn,usenatbib]{mnras}
\usepackage{newtxtext,newtxmath}
\usepackage[T1]{fontenc}
\usepackage{graphicx}	
\usepackage{amsmath}	

\def\deg{^\circ}

\def\kms{{\rm\,km\,s^{-1}}}
\def\msun{{\rm\,M_\odot}}

\def\eg{{ e.g.,\ }}

\def\lta{\mathrel{\spose{\lower 3pt\hbox{$\mathchar"218$}}
     \raise 2.0pt\hbox{$\mathchar"13C$}}}
\def\gta{\mathrel{\spose{\lower 3pt\hbox{$\mathchar"218$}}
     \raise 2.0pt\hbox{$\mathchar"13E$}}}

\def\FeH{{\rm[Fe/H]}}
\def\ione{\,{\sc i}}
\def\ii{\,{\sc ii}}

\newcommand{\mygi}{MyGIsFOS}
\newcommand{\teff}{\ensuremath{T_\mathrm{eff}}}
\newcommand{\glog}{\ensuremath{\log g}}
\newcommand{\eps}[1]{\log\varepsilon_{\rm #1}}

\title[C-19]{The \emph{Pristine}
survey -- XVII. The C-19 stream is dynamically hot and more extended than previously thought}

\author[Z. Yuan et al.]{
Zhen Yuan,$^{1}$\thanks{E-mail: zhen.yuan@astro.unistra.fr}
Nicolas F. Martin$^{1,2}$,
Rodrigo A. Ibata$^{1}$,
Elisabetta Caffau$^{3}$,
Piercarlo Bonifacio$^{3}$,
\newauthor
Lyudmila I. Mashonkina$^{4}$,
Rapha\"{e}l Errani$^{1}$,
Amandine Doliva-Dolinsky$^{1}$,
Else Starkenburg$^{5}$,
\newauthor
Kim A. Venn$^{6}$,
Anke Arentsen$^{1}$,
David S. Aguado$^{7}$,
Michele Bellazzini$^{8}$,
Benoit Famaey$^{1}$,
\newauthor
Morgan Fouesneau$^{2}$,
Jonay I. González Hernández$^{9,10}$,
Pascale Jablonka$^{11,3}$,
Carmela Lardo$^{12}$,
\newauthor
Khyati Malhan$^{2}$,
Julio F. Navarro$^{6}$,
Rubén Sánchez Janssen$^{13}$,
Federico Sestito$^{6}$,
\newauthor
Guillaume F. Thomas$^{9,10}$,
Akshara Viswanathan$^{5}$,
Sara Vitali$^{14}$
\\
$^{1}$Universit\'e de Strasbourg, CNRS, Observatoire Astronomique de Strasbourg, UMR 7550, F-67000 Strasbourg, France\\
$^{2}$Max-Planck-Institut f\"ur Astronomie, K\"onigstuhl 17, D-69117, Heidelberg, Germany \\
$^{3}$GEPI,Observatoire de Paris, Universit\'e PSL, CNRS, 5 Place Jules Janssen, 92190 Meudon, France \\
$^{4}$Institute of Astronomy, Russian Academy of Sciences, RU- 119017 Moscow, Russia\\ 
$^{5}$Kapteyn Astronomical Institute, University of Groningen, Landleven 12, NL-9747AD Groningen, the Netherlands \\
$^{6}$Department of Physics and Astronomy, University of Victoria, PO Box 3055, STN CSC, Victoria BC V8W 3P6, Canada \\
$^{7}$Dipartimento di Fisica e Astrofisica, Univerisit\'a degli Studi di Firenze, via G. Sansone 1, Sesto Fiorentino 50019, Italy\\
$^{8}$INAF-Osservatorio di Astrofisica e Scienza  dello  Spazio, via Gobetti 93/3, 40129 Bologna, Italy\\
$^{9}$Instituto de Astrofisica de Canarias, V\'{i}a L\'actea, 38205 La Laguna, Tenerife, Spain \\
$^{10}$Universidad de La Laguna, Departamento de Astrof\'isica, 38206 La Laguna, Tenerife, Spain \\
$^{11}$Laboratoire d'astrophysique, \'{E}cole Polytechnique F\'{e}d\'{e}rale de Lausanne (EPFL), Observatoire, 1290 Versoix, Switzerland \\
$^{12}$Dipartimento di Fisica e Astronomia, Universit\'{a} degli Studi di Bologna, Via Gobetti 93/2, I-40129 Bologna, Italy \\
$^{13}$UK Astronomy Technology Centre, Royal Observatory, Blackford Hill, Edinburgh, EH9 3HJ, UK \\
$^{14}$Núcleo de Astronomía, Facultad de Ingeniería y Ciencias Universidad Diego Portales, Ejército 441, Santiago, Chile
}

\date{Accepted XXX. Received YYY; in original form ZZZ}

\pubyear{}

\begin{document}
\label{firstpage}
\pagerange{\pageref{firstpage}--\pageref{lastpage}}
\maketitle

\begin{abstract}
The C-19 stream is the most metal poor stellar system ever discovered, with a mean metallicity $\FeH = -3.38\pm0.06$. Its low metallicity dispersion ($\sigma_{\rm [Fe/H]}$ $<$ 0.18 at the 95\% confidence level) as well as variations in sodium abundances strongly suggest a globular cluster origin. In this work, we use VLT/UVES spectra of seven C-19 stars to derive more precise velocity measurements for member stars, and to identify two new members with radial velocities and metallicities consistent with the stream's properties. One of these new member stars is located 30$\deg$ away from the previously identified body of C-19, implying that the stream is significantly more extended than previously known and that more members likely await discovery. In the main part of C-19, we measure a radial velocity dispersion $\sigma_v$ = 6.2$^{+2.0}_{-1.4}\kms$ from nine members, and a stream width of 0.56$\deg\pm0.08\deg$, equivalent to $\sim$158 pc at a heliocentric distance of 18 kpc. These confirm that C-19 is comparatively hotter, dynamically, than other known globular cluster streams and shares the properties of faint dwarf galaxy streams. On the other hand, the  variations in the Na abundances of the three newly observed bright member stars, the variations in Mg and Al for two of them, and the normal Ba abundance of the one star where it can be measured provide further evidence for a globular cluster origin. The tension between the dynamical and chemical properties of C-19 suggests that its progenitor experienced a complex birth environment or disruption history.

\end{abstract}


\section{Introduction}
\label{sec:intro}

Combining the Gaia Data Release EDR3 with the Pristine survey dedicated to the search for low-metallicity stars \citep{brown21, starkenburg17} we now have a sizable sample of stars with metallicity below [Fe/H] = $-2$ and accurate astrometric measurements \citep{lindegren21}. At the same time, continuous improvements in the search for substructures in the halo of the Milky Way have revealed a plethora of very metal-poor streams \citep{wan20, ibata21, li21b,martin22b}. Among them, the C-19 stream is the most metal-poor stream ever discovered. Its extreme metal-poor nature was first revealed by the photometric metallicities of member stars in the \emph{Pristine} survey, and further confirmed by spectroscopic follow-up studies \citep{martin22a}. Tight constraints on its unresolved metallicity dispersion and large variations in sodium abundances of its member stars strongly suggest that the C-19 stream is the remnant of a disrupted globular cluster (GC), whose extremely low metallicity is significantly below what measured for the lowest metallicity GCs in the Local Universe.


At the same time, the velocity dispersion of the C-19 stream is $\sim7\kms$ based on three neighboring members with observational uncertainty of $\sim2\kms$ and a rough estimate of the stream's end-to-end width gives $\sim$ 600 pc \citep{martin22a}. Judging from these measurements, the C-19 stream is dynamically hotter than most of the known globular cluster streams \citep[see \eg][and references within]{li21b}. In addition, \citet{martin22a} estimate the mass of the C-19 progenitor to be $\gtrsim10^4\msun$. This lower limit is based on the assumption that the full extent of the stream has been discovered. However, the possibility remains that C-19 is spatially more extended, although few obvious members (and no bound GC remnant) have so far been identified.

In this work, we confirm the unusually large velocity dispersion of the C-19 stream with accurate radial velocity measurements of confirmed members. We also explore the true extent of the C-19 stream through the discovery of a new member far away from the currently known part of the stream. We describe the target selection and observations in Sec.~\ref{sec:obs}. We present our results for the newly confirmed members and revised stream properties in Sec.~\ref{sec:res}. Finally, we conclude in Sec.~\ref{sec:con}.

\section{Observations and Spectral Analysis}
\label{sec:obs}

The list of all confirmed C-19 stars is provided in Table~\ref{tab:mem} and includes two new members (Pristine\_354.96$+$28.47 and C-19$\_$346.20$-$10.70) presented in this work for the first time. Pristine\_354.96$+$28.47 is a bright star ($G$ = 14.28) with very low metallicity (as indicated by photometric data from \emph{Pristine}) and that we were not able to observe previously. We observed this star on 18--21 May 2021 using VLT/UVES via program 0105.B-0235(A) (PI Ibata) with an exposure time of 600\,s.

In addition, we targeted potential members away from the main body of C-19 by using the \texttt{STREAMFINDER} to search for possible stream stars with a lower significance ($7\sigma$ instead of $10\sigma$, as defined by \texttt{STREAMFINDER}), located along the C-19 orbit derived in \citet{martin22a}, and with consistent proper motion measurements. Limiting the sample to $G < 17.5$ yields two potential member stars, located at $(\alpha,\delta)=(344.1263\deg,-17.0775\deg)$ and $(346.2032\deg,-10.6965\deg)$ that we also observed with VLT/UVES on 23--24 October 2021 via program 0108.B-0431(A) (PI Yuan). These two stars were initially observed with short 600\,s exposures and, after confirmation that C-19$\_$346.20$-$10.70 was a likely member based on its radial velocity, it was reobserved for an additional 1,800\,s for a total exposure time of 2,400\,s. The discrepant radial velocity of the other star clearly showed it is not a member of C-19 and it is not mentioned further in this paper.

Finally, we reobserved with VLT/UVES all five confirmed members listed in \citet{martin22a} that did not already have high-resolution spectroscopic follow-up and suffered from large radial velocity uncertainties. We observed the brightest of these stars (Pristine 355.32$+$27.59) for 1,800\,s and the other four for shorter, 600\,s exposures. In general, bright stars were observed at higher signal-to-noise to allow for chemical abundance determinations.

\begin{figure}
    \centering
    \includegraphics[width=\linewidth]{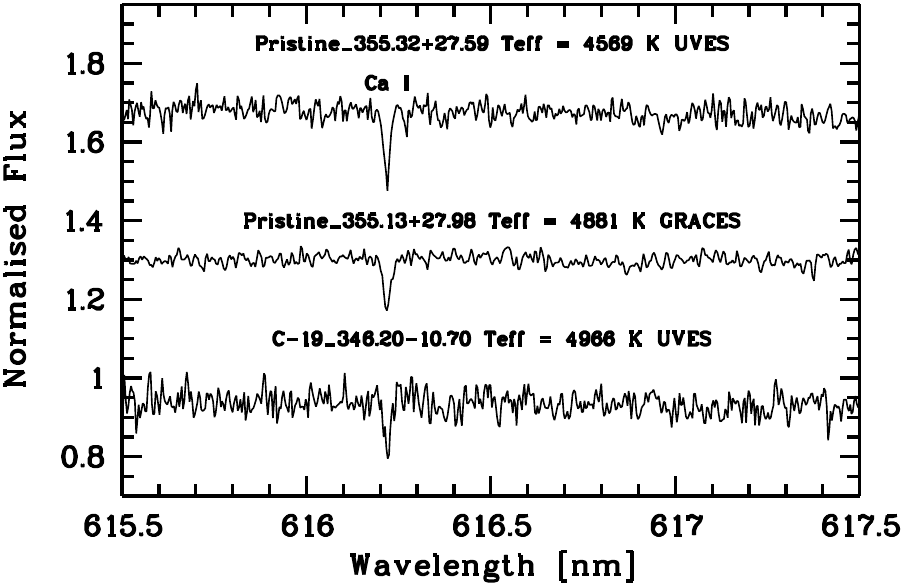}
    \caption{Portion of two of the newly acquired UVES spectra, compared to one of the GRACES spectra analysed in \citet{martin22a}.}
    \label{fig:obs}
\end{figure}

The setup of the UVES spectrograph was identical for these two programs, with the DIC2 dichroic beamsplitter in the wavelength range ``437+760", covering 3730–4990 \AA$~$ and 5650–9460 \AA. We used the 2 × 2 pixel binning readout mode and a $1"$ slit, yielding a resolution of R $\sim$ 40,000, to increase the efficiency for the fairly short exposures of both programs. The spectra were extracted and wavelength-calibrated using the \texttt{esoreflex} pipeline\footnote{https://www.eso.org/sci/software/esoreflex/}. In Fig.\,\ref{fig:obs}, a small portion of the reduced UVES spectra for Pristine$\_$355.13$+$27.98 and C-19$\_$346.20$-$10.70 is compared to the spectrum from the previous GRACES observations of \citet{martin22a}.
 
Radial velocities of all UVES spectra are measured through their cross-correlation against the spectrum of the radial velocity standard star HD\,182572, using the \texttt{fxcor} algorithm in IRAF.

We derive stellar parameters for all observed stars from Gaia EDR3 photometry \citep{riello21} and the \texttt{STREAMFINDER} distance estimate, as described in \citet{martin22a}, except for the effective temperature that was determined using theoretical colours rather than the \citet{MBM} Infra Red Flux Method calibration. These two temperature estimates differ by less than 80\,K, well inside the estimated uncertainties of  $\sim100$\,K. Chemical abundances are derived using the \mygi\ code \citep{sbordone14}, for which we use a grid of synthetic spectra computed under the assumption of local thermodynamical equilibrium (LTE) with SYNTHE from ATLAS 12 1D model atmospheres in hydrostatic equilibrium \citep{K05}. The atomic data used in computing the grid was taken from \citet{2021A&A...645A.106H} for molecules we used the data 
available from the site of R.L. Kurucz \footnote{\url{}http://kurucz.harvard.edu/molecules.html} and for CH
we used the line list of \citet{masseron}.
\mygi\ provides local fits to pre-identified features, that may include several atomic and molecular data. We have successfully used \mygi\ in other papers of the series \citep{paperII,paperV,paperXI} and,
in \citet{paperIV}, we successfully benchmarked it
against other chemical analysis procedures. 
The carbon abundance was instead derived by direct fitting of the G-band, after computing synthetic spectra with several carbon abundances. We used the CH line list of \citet{masseron}.

We calculate the non-local thermodynamic equilibrium (NLTE) abundance
corrections by using the methods developed and tested in our earlier studies for Na\ione\ \citep{alexeeva_na}, Mg\ione\ \citep{mash_mg13}, Al\ione\ \citep{mash_al2016}, K\ione\ \citep{2020AstL...46..621N}, Ca\ione -\ii\ \citep{2017AA...605A..53M}, Ti\ione -\ii\ \citep{sitnova_ti}, Fe\ione -\ii\ \citep{mash_fe}, Sr\ii\ \citep{2022MNRAS.509.3626M}, and Ba\ii\ \citep{2019AstL...45..341M}. The atom models are comprehensive and implement the most up-to-date atomic data available for atomic energy levels, transition probabilities, photoionization cross sections, rate coefficients for excitation by collisions with electrons and hydrogen atoms, charge exchange $A + \rm{H^-}$ processes, and electron-impact ionization. The coupled radiation transfer and statistical equilibrium (SE) equations are solved with the code {\sc detail} \citep{detail}, where the opacity package was revised following \citet{mash_fe}. The LTE and NLTE level populations from {\sc detail} were then used by the code {\sc linec} \citep{Sakhibullin1983} that computes the NLTE abundance corrections, $\Delta_{\rm NLTE} = \eps{NLTE}-\eps{LTE}$, for individual spectral lines. 

\section{Results}
\label{sec:res}
\begin{figure*}
\centering
\includegraphics[width=\linewidth]{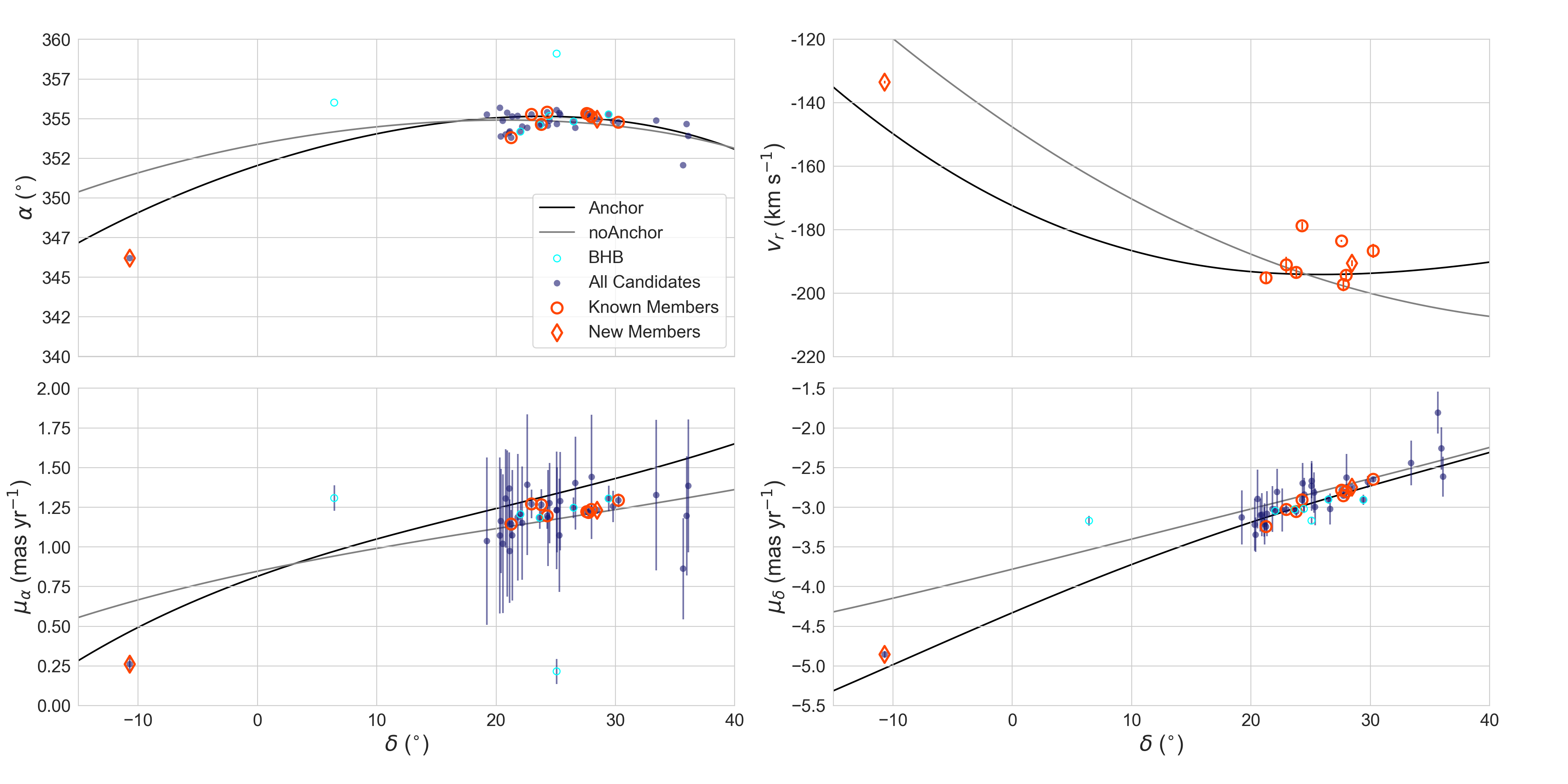}
\vspace*{-6mm}
\caption{The C-19 stream in 5-D phase space. All 38 candidate members in the main body of the C-19 stream from \citet{martin22a} are represented by solid blue circles, except for the possible BHB stars that are shown as open cyan circles. The eight known members with updated, accurate velocity measurements are represented by orange circles, and the two new members as orange diamonds. The orbit anchored at 18 kpc (black line) aligns closer to the new member located 30$\deg$ away from the main body of the stream, compared to the orbit without distance anchoring (grey line).}
\label{fig:c-19}
\end{figure*}

\subsection{Revised velocities and new members}
\label{subsec:mem}

In Fig.~\ref{fig:c-19}, we show the distribution of likely C-19 members \textbf{solid blue circles} on the sky and in velocity space. The black and grey lines correspond to the orbits presented by \citet{martin22a} for C-19, either anchored at the favored distance of 18 kpc (the average distance of the BHB members), or based on the distance of individual stars predicted by \texttt{STREAMFINDER}. Members previously confirmed through radial velocities are highlight as open orange circles, and the two new members confirmed in this study are shown as open orange diamonds. All the BHB stars are highlighted as open cyan circles, and three of them are excluded from the candidate list because they are relatively off-track in sky positions or in proper motion space. The radial velocities precisions measured from the UVES spectra are significantly improved for the 5 members that were only observed at low resolution with GTC/OSIRIS in \citet{martin22a}, with average uncertainties updated from $\sim15\kms$ to 1--2$\kms$. For the two new members, Pristine$\_$354.96$+$28.47 is in the main body of the stream, but was not identified before because it was not in the \emph{Pristine} catalog used in \citet{martin22a}\footnote{When the spectroscopic observations of \citet{martin22a} were designed, the catalog of Pristine photometric metallicities relied on SDSS-broadband photometry (see \citealt{starkenburg17}) and this bright star was saturated in the SDSS. It therefore never made it into the sample of candidate extremely metal-poor stars that were likely C-19 members. Since then, we developed a version of the Pristine photometric metallicity catalog that instead uses Gaia broadband photometry and easily revealed Pristine\_354.96+28.47 (G=14.28) as an extremely metal-poor candidate and a very likely C-19 member.}. In this work, we show that it has a radial velocity that is in very good agreement with those of the other confirmed members. C-19$\_$346.20$-$10.70 is located $\sim$ 30$^{\circ}$ away but has a radial velocity that is close to the velocity of the calculated C-19 orbit at this location on the sky. We also show below that it has the same extremely low metallicity as other C-19 members (see below). The anchored orbit is in good agreement with the 5D kinematics of C-19$\_$346.20$-$10.70, hinting that there could be more potential members spread along this orbit between the main body of C-19 and this new member. 

Based on our new UVES spectrum, we revise the metallicity measurement of Pristine$\_$355.32$+$27.59. We derive [FeI/H]$_{\rm LTE}$ = $-$3.49$\pm$0.21, [FeI/H]$_{\rm NLTE}$ = $-$3.29$\pm$0.21, and [FeII/H] = $-$3.55$\pm$0.26, which deviates slightly from but is nevertheless compatible with the previous measurement based on a GTC/OSIRIS spectrum ([Fe/H]$_{\rm LTE}$ = $-$3.15$\pm$0.18; \citealt{martin22a}). This measurement is also in perfect agreement with the previously reported mean metallicity of C-19 ($\FeH=-3.38\pm0.06$; \citealt{martin22a}). The metallicity of the two new members is also in agreement with this average metallicity: the spectrum of C-19$\_$346.20$-$10.70 yields [FeI/H]$_{\rm LTE}$=$-$3.44$\pm$0.16, [FeI/H]$_{\rm NLTE}$=$-$3.31$\pm$0.16, and [FeII/H]=$-$3.26$\pm$0.18 while, from the lower signal-to-noise spectrum of Pristine$\_$354.96$+$28.47, we are able to derive [FeI/H]$_{\rm LTE}$=$-$3.54$\pm$0.11, [FeI/H]$_{\rm NLTE}$=$-$3.28$\pm$0.11. Based on the radial velocity and metallicity measurements, we are confident these two stars are C-19 members.

Table\,\ref{tab:mem} lists all the metallicity measurements from spectroscopic follow-up studies based on LTE assumptions. Note that [FeI/H]$_{\rm LTE}$ and  [FeII/H] (which is free of the NLTE effects) are inferred from both the Gemini/GRACES and VLT/UVES spectra. [Fe/H]$_{\rm LTE}$ are derived for GTC/OSIRIS spectra.

\subsection{Chemical Properties}
\label{subsec:chem}

\begin{figure*}
    \centering
    \resizebox{15truecm}{!}{
    \includegraphics{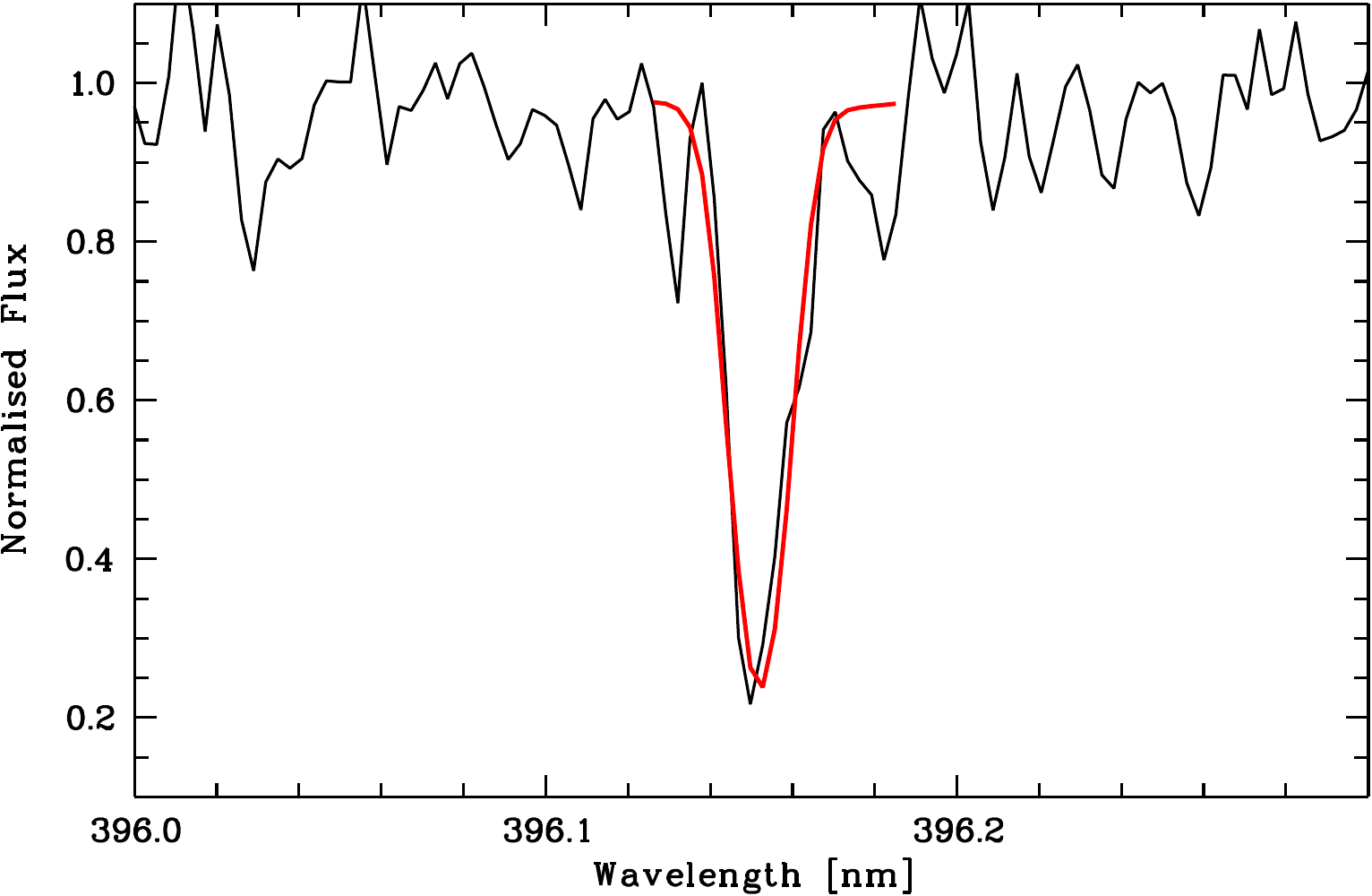}
    \hspace{1cm}
    \includegraphics{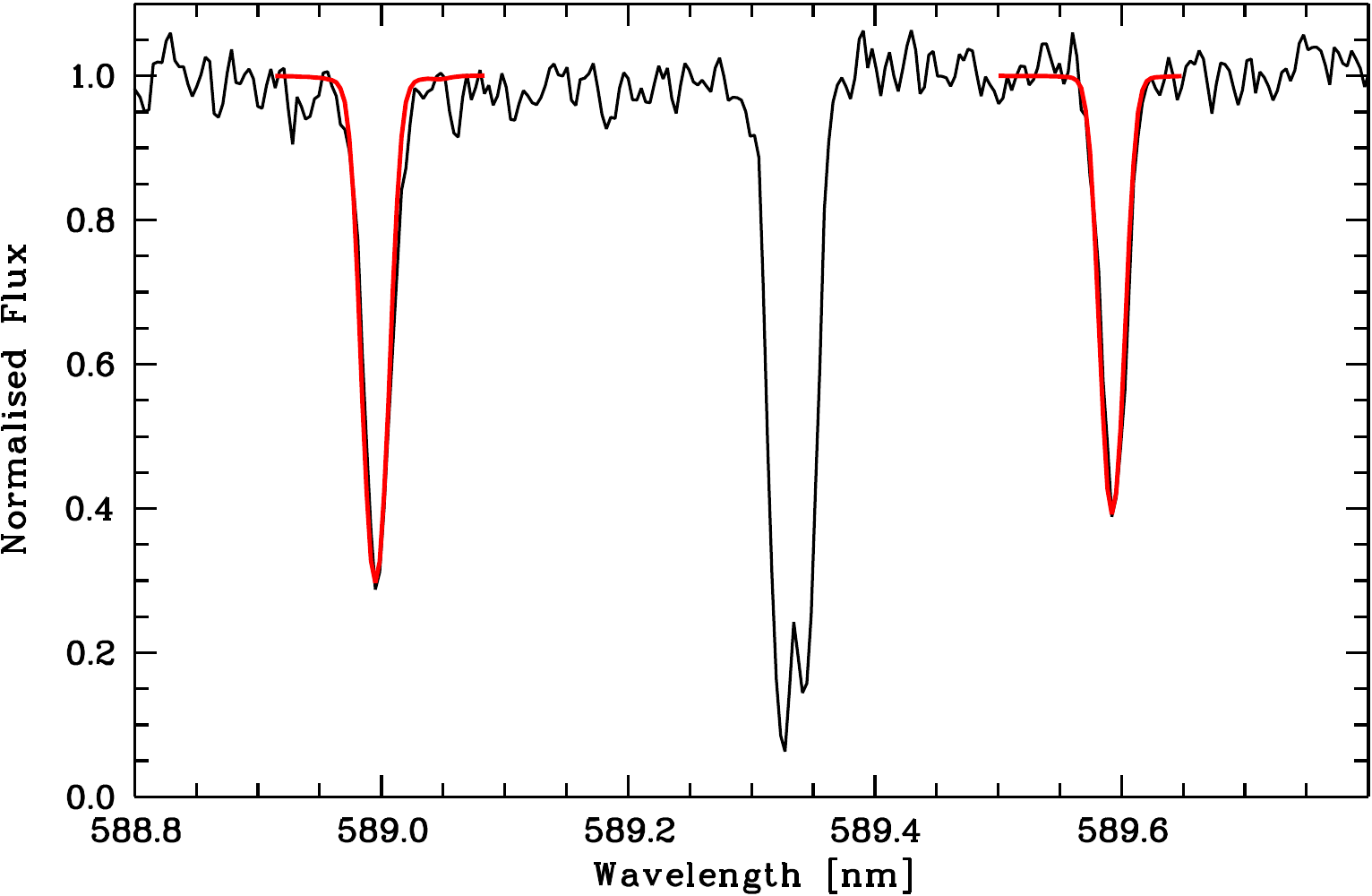}
    }
    \caption{Left panel: The Al\ione\ resonance line at 396.1\,nm for star Pristine$\_$355.32$+$27.59, along with the best LTE fit by \mygi (red line). Right panel: The same for the Na\ione D doublet. The strong absorption between the two stellar NaD lines is the interstellar D2 line.}
    \label{fig:Al_Na_355}
\end{figure*}

\begin{figure*}
\centering
\includegraphics[width=\linewidth]{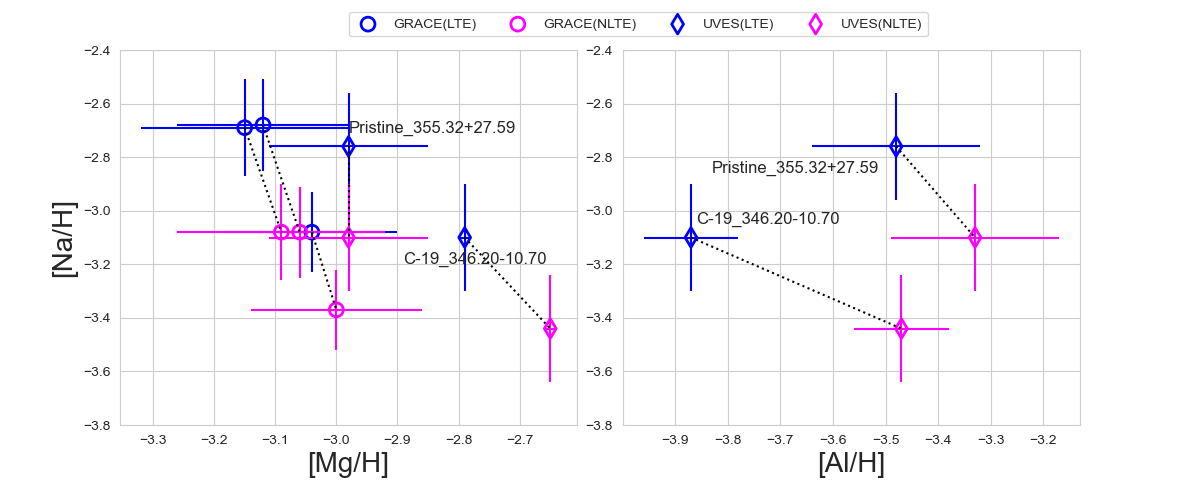}
\caption{Distribution of [Na/H], [Mg/H], and [Al/H] for the C-19 stream stars investigated in this study (diamonds) and by \citet[][circles]{martin22a}. The NLTE and LTE data are represented by the magenta and blue colour, respectively, and linked for the same star. The stars show variations in the three elements, whether one considers LTE or NLTE measurements.}
\label{fig:elem}
\end{figure*}

For three of the stars observed with UVES, including the two new members, we determine abundances for other elements besides iron. These are provided in Table~\ref{tab:abun}, together with the adopted atmospheric parameters. $A(X)_\odot$ was used to derive [X/H]. The table provides the line-to-line scatter for elements where more than one line is measured and the
average NLTE corrections for the measured lines
of several chemical species (Na\ione, Mg\ione, Al\ione, K\ione, Ca\ione, Ti\ione -\ii, Fe\ione -\ii, Sr\ii, Ba\ii). Fig.\,\ref{fig:Al_Na_355} also shows two portions of the spectrum of Pristine$\_$355.32$+$27.59, around the the Al\ione\ resonance line at 396.1\,nm and the Na\ione D doublet, along with the best LTE fit for those lines.

Both NLTE and LTE chemical abundances of the three newly observed stars strengthen the scenario that the C-19 stream is the remnant of a globular cluster. Two stars have a Na abundance that is 0.3\,dex higher than the third one (C-19$\_$346.20$-$10.70), consistent with the notion that the stars with higher Na are second generation stars \citep[see e.g.][and references therein]{2004ARA&A..42..385G,2018ARA&A..56...83B} and consistent with what was found by \citet{martin22a}. 
Pristine$\_$355.32+27.59 has [Mg/H] = $-2.98$, which is, within the uncertainties, consistent with the Mg abundances of the three C-19 stars from \citet{martin22a}. This means that using different Mg\ione\ lines (3832, 4571, 8806\,\AA) than \citet[][Mg\ione\ 5172, 5183, 5528\,\AA]{martin22a} does not lead to systematic abundance shifts. C-19$\_$346.20$-$10.70 has the highest with [Mg/H] = $-2.65$, which deviates from the Mg abundance in the other stars and hint at a possiblen intrinsic variation of Mg in the cluster. This star also has the lowest [Na/H] (Fig.\,\ref{fig:elem}).
Finally, 
Pristine$\_$355.32$+$27.59 has a higher Al abundance than C-19$\_$346.20$-$10.70 and provides further evidence that the material
has undergone Mg-Al cycling, besides Ne-Na cycling. The S/N ratio of our spectra is insufficient to derive O abundances, which would be needed to confirm the Na-O anti-correlation that is typically found in globular clusters.

\begin{figure}
    \centering
   \resizebox{7.5cm}{!}{\includegraphics{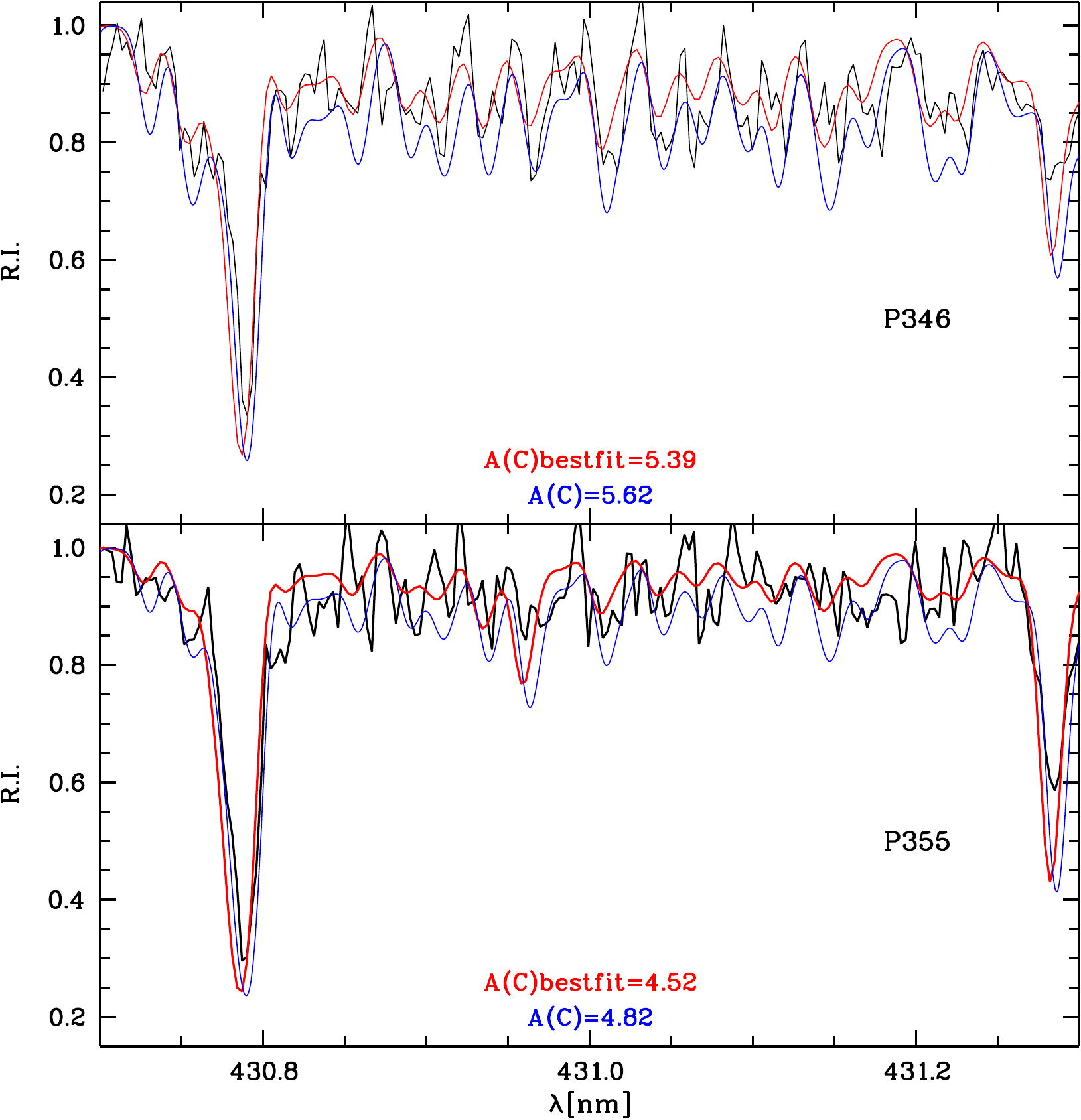}}
    \caption{Portion of the G-band spectrum for Pristine$\_$355.32$+$27.59 and C-19$\_$346.20$-$10.70 together with the best fitting spectrum and a spectrum with A(C) 0.3\,dex lower, i.e. $1\sigma$ lower.}
    \label{fig:C346}
\end{figure}

For Pristine$\_$355.32$+$27.59 and C-19$\_$346.20$-$10.70 we were able to measure the carbon abundances from the G-band as shown in Fig.~\ref{fig:C346}. Pristine$\_$355.32$+$27.59 shows a depleted carbon abundance, as expected for a star of this luminosity that has mixed material processed through the CNO cycle. We expect nitrogen to be enhanced in this star. C-19$\_$346.20$-$10.70 has a slightly enhanced C abundance ([C/FeII] = $+$0.3), as often observed in field stars of this metallicity \citep[see e.g][]{spite05}.

\begin{table*}
	\centering
	\caption{C-19 Members with Spectroscopic Follow-ups}
	\label{tab:mem}
	\begin{tabular}{cccccccccc} 
		\hline
	Name & RA & DEC & G & t$_{\rm exp}$ & $v_{\rm r}$ &  [FeI/H]$_{\rm LTE}$ &  [FeII/H] &[Fe/H]$_{\rm LTE}$ & Instrument\\
	 & (deg) & (deg) && (s) & (km s$^{-1}$) & HR & HR & OSIRIS\\
		\hline
Pristine$\_$354.77$+$30.25&354.7701&$+$30.2509&16.33&2400&$-$186.7$\pm$2.2&$-$3.21$\pm$0.17&$-$3.42$\pm$0.12&&GRACES \\
Pristine$\_$355.27$+$27.74&355.2755&$+$27.7483&16.51&4800&$-$197.3$\pm$2.1&$-$3.15$\pm$0.17&$-$3.42$\pm$0.17&&GRACES \\
Pristine$\_$355.13$+$27.98&355.1327&$+$27.9819&16.16&7200&$-$194.4$\pm$2.0&$-$3.30$\pm$0.15&$-$3.45$\pm$0.11&$-$3.32$\pm$0.18&OSIRIS \& GRACES \\
Pristine$\_$353.79$+$21.26 &353.7995&$+$21.2648&17.63&600&$-$195.2$\pm$1.9&&&$-$3.41$\pm$0.19& OSIRIS \& UVES \\
Pristine$\_$354.63$+$23.79 &354.6358&$+$23.7941&17.49&600&$-$193.5$\pm$1.6&&&$-$3.38$\pm$0.20& OSIRIS \& UVES \\
Pristine$\_$355.26$+$22.96 &355.2671&$+$22.9677&17.60&600&$-$191.1$\pm$2.6&&&$-$3.34$\pm$0.19&OSIRIS \& UVES \\
Pristine$\_$355.32$+$27.59 &355.3223&$+$27.5993&14.19&1800&$-$183.6$\pm$0.3&$-$3.49$\pm$0.21& $-$3.55$\pm$0.26 &$-$3.15$\pm$0.18&OSIRIS \& UVES \\
Pristine$\_$355.40$+$24.28 &355.4056&$+$24.2832&17.28&600&$-$178.8$\pm$1.2&&&$-$3.39$\pm$0.18&OSIRIS \& UVES \\
$^{\dagger}$Pristine$\_$354.96$+$28.47 &354.9615&$+$28.4659&14.28&600&$-$190.6$\pm$0.9&$-$3.48$\pm$0.11&&&UVES \\
$^{\dagger}$C-19$\_$346.20$-$10.70 &346.2032&$-$10.6965&15.89&1800&$-$133.5$\pm$0.4&$-$3.51$\pm$0.19 & $-$3.41$\pm$0.18&&UVES \\
		\hline
	\end{tabular}
Note: [FeI/H]$_{\rm LTE}$ and [FeII/H] (unaffected by NLTE corrections) are determined from both the Gemini/GRACES and VLT/UVES spectra. [Fe/H]$_{\rm LTE}$ is derived for the GTC/OSIRIS spectra. Starting from the forth row, seven members with UVES spectra in this work are listed, and the last two stars with $^{\dagger}$ are new confirmed members.
\end{table*}

\begin{table*}
    \centering
    \caption{Atmospheric parameters and chemical abundances for three of the targets
observed with UVES.}
    \begin{tabular}{lrrrrrrrrrrrrrc}
    \hline
         &   \multicolumn{4}{c}{Pristine$\_$355.32$+$27.59} & 
        \multicolumn{4}{c}{Pristine$\_$354.96$+$28.47}&
        \multicolumn{4}{c}{C-19$\_$346.20$-$10.70}
       \\ 
    \hline         
         \teff & 4569\,K & & & &  4446\,K& & & &4966\,K \\ 
\glog          &   0.95 & &  & &      0.87& & &  &2.06\\             
$vturb$        &   2.21 & & & &       2.19& & && 1.88   \\ 
S/N@ 600\,nm & 32      & & & &       9   & & && 10 \\
\hline\\
ion            &  [X/H] & $\sigma$& $N$ & NLTE$^{\dagger}_c$ & [X/H] & $\sigma$& $N$ & NLTE$^{\dagger}_c$& [X/H] & $\sigma$& $N$ & NLTE$^{\dagger}_c$& $A^{\ast}_\odot$\\
\hline\\
\relax [FeI/H]        & $-3.49$&$0.21$ &34&$+0.2$ &$-3.48$&0.10&3& $+0.2$&  $-3.51$&0.19&33&$+0.2$  & 7.52 \\
\relax [FeII/H]       & $-3.55$&$0.26$ &3 & 0.0   &       &    & &       &  $-3.41$&    &1 &$0.0$   & 7.52 \\
\relax [C/H]          &   $-3.98$ & 0.3 & G-band & & & & & & $-3.11$ & 0.3 &G-band & & 8.50\\
\relax [Na/H]         & $-2.76$&$0.20$ &2 &$-0.34$&$-2.80$&0.13&2& $-0.31$ &  $-3.10$&   &1&$-0.34$ & 5.30 \\
\relax [Mg/H]         & $-2.98$&$0.13$ &3 &$0.0$&       &    & &         &  $-2.79$&0.01&2&$+0.14$& 7.54 \\
\relax [Al/H]         & $-3.48$&$0.16$ &2 &$+0.15$&       &    & &         &  $-3.87$&0.09&2&$+0.4$ & 6.47 \\
\relax [K/H]          & $-2.85$&$    $ &1 &$-0.09$&       &    & &         &  $     $&    & &       & 5.11 \\
\relax [CaI/H]        & $-3.08$&$0.06$ &4 &$+0.2$ &       &    & &         &  $-2.99$&0.06&3&$+0.09$& 6.33 \\
\relax [ScII/H]       & $-3.50$&$0.09$ &2 &       &       &    & &         &  $-3.02$&    &1&       & 3.10\\
\relax [TiI/H]        & $-3.51$&$0.11$ &2 &$+0.4$ &       &    & &         &  $-2.95$&0.14&2&$+0.3$ & 4.90\\
\relax [TiII/H]       & $-3.14$&$0.15$ &14&$+0.05$&       &    & &         &  $-2.87$&0.23&7&$+0.05$& 4.90\\
\relax [MnI/H]        & $-4.49$&$    $ &1 &       &       &    & &         &  $-4.34$&    &1&       & 5.37\\
\relax [Co/H]         & $-3.51$&$0.17$ &7 &       &       &    & &         &$-3.09$&0.20&4&         & 4.92\\
\relax [Ni/H]         & $-3.86$&$0.17$ &3 &       &       &    & &         &$-3.32$&0.27&3&         & 6.23\\
\relax [Sr/H]         & $-3.86$&$0.19$ &3 &$+0.07$&       &    & &         &$-3.80$&0.01&2&$+0.03$   & 2.92\\
\relax [Y/H]          & $-4.11$&$    $ &1 &       &       &    & &         &       &    & &         &2.21\\
\relax [Ba/H]         & $-4.28$&$    $ &1 &$+0.09$ &       &    & &         &       &    & &         &2.17\\
\hline
\multicolumn{13}{l}{$^{\dagger}$ The NLTE corrections, NLTE$_c$ provided in the table are the average of the NLTE corrections of the lines used for which we could compute} \\
\multicolumn{13}{l}{the statistical equilibrium.}\\
\multicolumn{13}{l}{$^{\ast}$ The solar chemical abundances are taken from \citet{lodders09,caffau11}.} \\
    \end{tabular}
    \label{tab:abun}
\end{table*}

\subsection{Dynamical Properties}
\label{subsec:dyn}

\begin{figure}
\centering
\includegraphics[width=\linewidth]{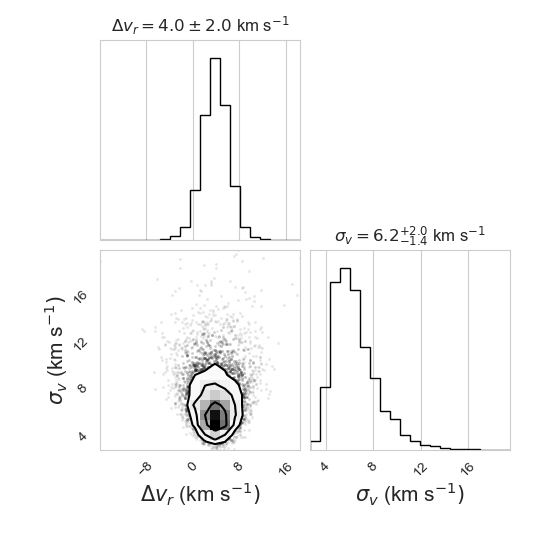}
\vspace*{-6mm}
\caption{PDFs of the radial velocity offset from the favored orbit ($\Delta v_r$) and the radial velocity dispersion ($\sigma_v$), inferred from the sample of 9 member stars in the main body of C-19. of the sample. The bottom-left panel shows the two-dimensional PDF, taken directly from the Markov Chain Monte Carlo chain and the other two panels display the marginalized one-dimensional PDFs for the two parameters.}
\label{fig:corner}
\end{figure} 

We derive the velocity dispersion of the main part of the C-19 stream from all stars with accurate radial velocities obtained from GRACES and UES high-resolution spectra. Since the radial velocity of the favored anchored C-19 orbit does not vary significantly with declination $\delta$ in the main part of the stream (see upper-right panel of Fig.~\ref{fig:c-19}), we infer the mean radial velocity offset from the orbit and the corresponding velocity dispersion by assuming that the distribution of $\Delta v_r(\delta)=v_r-v_{r,\rm orbit}(\delta)$, where $v_{r,\rm orbit}(\delta)$ is the radial velocity of the anchored orbit for declination $\delta$, follows a Gaussian distribution around the orbit. From the 9 stars located in the main body of C-19, we use the formalism of \citet{martin18}, to infer the mean offset of the sample of velocities from the orbit, $\Delta v_r$ = 4.0$^{+2.1}_{-2.3}\kms$, and its dispersion, $\sigma_v$ = 6.2$^{+2.0}_{-1.4}\kms$. The resulting probability distribution functions (PDFs) for those two parameters are shown in Fig.~\ref{fig:corner}. We also derive the average orbital plane of C-19 from these stars, yielding 87$\deg$ with respect to the Galactic plane.

To constrain the width of the stream, we apply the same approach to the 38 candidate members in the main body of the stream that are shown as blue circles in Fig.~\ref{fig:c-19}. We infer a Gaussian stream width of 0.56$\deg\pm0.08\deg$, which translates to $\sim$159 pc with the assumed anchoring distance of 18 kpc.

Both the velocity dispersion of the stream and its width show that the C-19 stream is dynamically hotter than most of the globular cluster streams. These all have velocity dispersions smaller than $5\kms$, often significantly so \citep[see \eg][and references within]{li21b}. For example, the GD-1 stream has a velocity dispersion $\sigma_v$ = 2.1$\pm$0.3$\kms$ \citep{gialluca21}, and approximate stream width of 0.5$\deg$, or $\sim$100 pc at 10 kpc \citep{koposov10, malhan18b}. Similarly, the Palomar 5 tidal stream has $\sigma_v$ = 3.2$\kms$ \citep{ishigaki16}. While the chemical abundances of C-19 stars indicate that its progenitor was a globular cluster, a companion study shows that the combined large width and velocity dispersion prove difficult to reconcile with the disruption of a simple globular cluster \citep{errani22}. Having the C-19 progenitor globular cluster be processed in its own dark matter subhalo before it is accreted onto the MW could alleviate this tension. Various numerical studies suggest that the width and velocity dispersion of globular cluster streams that have been pre-processed in a dark matter subhalo are larger than those of GCs without hosts \citep{penarrubia17, carlberg18, malhan21a, carlberg20, carlberg21, errani22}. Such configuration can delay the stream formation and could explain why C-19 is still coherent today \citep{vitral21}. Alternatively, tidal heating from dark matter subhalos \citep[see \eg][]{Ibata2002heating,Johnston2002,Penarrubia2019} or giant molecular clouds \citep[see \eg][]{amorisco16} might increase the width and dispersion of the stream. For example, the ATLAS-Aliqa Uma stream has an exceptionally large velocity dispersion of $\sim$ 6$\kms$ around the discontinuity region of the stream (its ``kink'' feature), which is considered as the sign of perturbations \citep{li21a}. Since the velocity dispersion of a stream reflects the properties of its progenitor, as well as the encounters along its disruption history, it is difficult to directly use it as an indicator of the stream's origin.

The newly confirmed member, C-19$\_$346.20$-$10.70, located more than $30\deg$ away from the main body of C-19, implies that additional C-19 members could be found along the stream's orbit. It ensues that the mass estimate of C-19 calculated by \citet[$\gtrsim 0.8 \times 10^{4}\msun$]{martin22a} is indeed a lower limit to the total mass of its progenitor. However, as pointed out in that study, the progenitor is unlikely to be much more massive than $10^{4}\msun$, given the upper limit of $10^{4.65}\msun$ suggested by the galaxy mass-metallicity relation from zoom-in cosmological simulations \citep{ma16, kruijssen19}.

\section{Conclusions}
\label{sec:con}

In this work, we have derived accurate radial velocities for seven stars that belong to the extremely metal-poor C-19 stream and have measured detailed chemical abundances for three of them, using VLT/UVES spectra. Based on these results, we confirm two new C-19 members, one of which is located more than 30$\deg$ away from the previously discovered main body of the stream. The three members with derived chemical abundances all have extremely low metallicities, in good agreement with the average metallicity of C-19 ($\FeH = -3.38\pm0.06$). They also exhibit variations in Na abundance that are larger than 0.3\,dex, and show the anti-correlation between Na and Mg and Al that is often found in globular clusters, further supporting the idea that the C-19 progenitor was a globular cluster.

Based on the 9 members with accurate radial velocities in the main body of the stream, we determine an intrinsic velocity dispersion of 6.2$^{+2.0}_{-1.4}\kms$ consistent with the previous estimate of $\sim7\kms$ based on only 3 member stars \citep{martin22a}. The velocity dispersion of C-19 is larger than that of other globular cluster streams with accurate measurements. Besides, the stream width of $\sim$158 pc is also on the large side of the currently observed range \citep[see \eg][]{li21b}. These refined properties of C-19 are used in a companion study by \citet{errani22} to show that the simple tidal disruption of a globular cluster on the C-19 orbit cannot naturally reproduce the stream. This tension could be alleviated if the C-19 progenitor globular cluster was pre-processed in its own dark matter halo or was heated when orbiting around the Milky Way, either from disk shocking, or via the nearby passage of dark matter subhalos or giant molecular clouds.


\section*{Data Availability}
\label{sec:data}
The C-19 memberlist is published via the Github repository with the link of \url{https://github.com/zyuan-astro/C-19}.

\section*{Acknowledgements}

Z.Y., NFM, and RAI acknowledge funding from the Agence Nationale de la Recherche (ANR project ANR-18-CE31-0017). Z.Y., NFM, RAI, RE, AA and BF also acknowledge funding from the European Research Council (ERC) under the European Unions Horizon 2020 research and innovation programme (grant agreement No. 834148). ES acknowledges funding through VIDI grant "Pushing Galactic Archaeology to its limits" (with project number VI.Vidi.193.093) which is funded by the Dutch Research Council (NWO). KAV is grateful for funding through the Canadian National Science and Engineering Research Council Discovery Grants and CREATE programs. DA acknowledges support from the ERC Starting Grant NEFERTITI H2020/808240. JIGH acknowledges financial support from the Spanish Ministry of Science and Innovation (MICINN) project PID2020-117493GB-I00. 

We gratefully acknowledge the High Performance Computing center of the Université de Strasbourg for a very generous time allocation and for their support over the development of this project.

This work has made use of data from the European Space Agency (ESA) mission {\it Gaia} (\url{https://www.cosmos.esa.int/gaia}), processed by the {\it Gaia} Data Processing and Analysis Consortium (DPAC, \url{https://www.cosmos.esa.int/web/gaia/dpac/consortium}). Funding for the DPAC has been provided by national institutions, in particular the institutions participating in the {\it Gaia} Multilateral Agreement.

Based on observations collected at the European Southern Observatory under ESO programmes 0105.B-0235(A) and 0108.B-0431(A).

Software: STREAMFINDER \citep{malhan18a}, cornerplot \citep{cornerplot}, astropy \citep{astropy}, galpy \citep{galpy}, IRAF \citep{tody86,tody93}, numpy \citet{numpy}, scipy \citep{scipy}, matplotlib \citep{matplotlib}, seaborn \citep{seaborn}, UltraNest\footnote{\url{https://johannesbuchner.github.io/UltraNest/}} \citep{buchner21}.

\bibliographystyle{mnras}
\bibliography{ms}


\bsp	
\label{lastpage}
\end{document}